\documentclass[aps,prd,reprint,superscriptaddress]{revtex4-1}
\usepackage{amsmath,amssymb}
\usepackage[colorlinks=true,allcolors=blue]{hyperref}
\usepackage{dcolumn}
\usepackage{units}
\usepackage{graphicx}
\usepackage{float}
\usepackage{enumitem}
\usepackage{diagbox}
\usepackage{overpic}
\usepackage{wasysym}
\usepackage{comment}
\usepackage{color}

\allowdisplaybreaks[1]

\newcommand{\s}{{\text{s}}}
\newcommand{\e}{{\text{e}}}
\newcommand{\m}{{\text{m}}}

\newcommand{\ve}[1]{\mathbf{#1}}
\newcommand{\be}{\begin{equation}}
\newcommand{\ee}{\end{equation}}
\newcommand{\eref}[1]{Eq.~(\ref{#1})}
\newcommand{\fref}[1]{Fig.~(\ref{#1})}
\newcommand{\tref}[1]{Table~\ref{#1}}
\newcommand{\sref}[1]{Sec.~\ref{#1}}

\begin{document}
\title{Testing the Strong Equivalence Principle with spacecraft ranging towards the 
nearby Lagrangian points}

\author{Giuseppe Congedo}
\email{giuseppe.congedo@physics.ox.ac.uk}
\affiliation{Department of Physics, University of Oxford, Keble Road, Oxford, OX1 3RH, United Kingdom}

\author{Fabrizio De Marchi}
\email{fabrizio.demarchi@uniroma1.it}
\affiliation{Dipartimento di Ingegneria Meccanica e Aerospaziale, Universit\`{a} degli Studi di Roma ``La Sapienza'', via Eudossiana, 18 -- I-00184 Roma, Italy}

\date{\today}

\begin{abstract}
General relativity is supported by great experimental evidence. 
Yet there is a lot of interest in precisely setting its limits with on going and future experiments.
A question to answer is about the validity of the Strong Equivalence Principle.
Ground experiments and Lunar Laser Ranging have provided the best upper limit on the Nordtvedt parameter $\sigma[\eta]=4.4\times 10^{-4}$.
With the future planetary mission BepiColombo, 
this parameter will be
further improved by at least an order of magnitude.
In this paper we envisage yet another possible testing environment with spacecraft
ranging towards the nearby Sun-Earth collinear Lagrangian 
points. 
Neglecting errors in planetary masses and ephemerides,
we forecast $\sigma[\eta]=6.4\,(2.0)\times10^{-4}$ (5 yr integration time) via ranging towards $L_1$ in a realistic (optimistic) scenario 
depending on current (future) range capabilities and knowledge of the Earth's ephemerides. 
A combined measurement, $L_1$+$L_2$, gives instead $4.8\,(1.7)\times10^{-4}$. 
In the optimistic scenario a single measurement of one year would be enough to reach $\approx3\times10^{-4}$. 
All figures are comparable with Lunar Laser Ranging, but worse than BepiColombo. Performances could be much improved if data were 
integrated over time and over the number of satellites flying around either of the two Lagrangian points.
We point out that some systematics (gravitational perturbations of other 
planets or figure effects) are much more in control compared to other experiments. 
We do not advocate a specific mission to constrain the Strong Equivalence Principle, 
but we do suggest analysing ranging data of present and future spacecrafts flying around $L_1$/$L_2$ (one key mission is, for instance, LISA Pathfinder).
This spacecraft ranging would be a new and complementary probe to constrain the Strong Equivalence Principle in space.
\end{abstract}


\maketitle

\section{Introduction}

General relativity 
provides the most satisfying physical description of gravity with a great experimental evidence \cite{will2014}.
The equivalence principle (EP) lies at the heart of general relativity and states the equivalence between inertial and gravitational mass.
According to it, the roles of inertial and gravitational mass can be mutually interchanged without affecting the observed dynamics of test masses. 
The universality of free fall is therefore a direct consequence of this principle.
A key observable in general relativity is the Riemann tensor, which describes the local gravity's tidal field between free falling test masses.
Evidently, a difference between inertial and gravitational mass shows up as a differential acceleration between free falling test masses. 
Much like gravitational wave detection where a differential acceleration is induced between free falling test masses \cite{congedo2015b,congedo2015a}, testing the equivalence principle requires measuring a differential acceleration that would not 
otherwise be present if general relativity were the correct and ultimate theory of gravity.
The weak form of the EP (WEP) can be verified with test masses of different chemical compositions. 
However the strong EP (SEP) extends the validity of the principle to self-gravitating bodies with different self-energies, and therefore is much harder to test. 
The WEP can in fact be tested on ground with, for instance, torsion balances \cite{adelberger2009} 
or in space with low-earth orbits (e.g.\ with the future MICROSCOPE mission \cite{touboul2012}), 
pushing the limits of the equivalence principle down to $\sigma [\delta a/a]\approx10^{-15}$. 
On the contrary, the SEP requires an experiment specifically devised in space with much longer baselines and bigger masses 
over distances of some AU \cite{turyshev2004}.
Even though Lunar Laser Ranging (LLR) can constrain both the WEP and SEP with remarkable results over the years \cite{murphy2012}, 
missions in the solar system, like for instance BepiColombo \cite{milani2002}, provide a better framework for the SEP as the involved self-energies are much bigger. 
In addition, the discovery of the triple system J0337+1715 \cite{ransom2014}, made of a pulsar and two white dwarves, has recently shed new light onto the concrete possibility of testing the SEP outside the solar system. 
The very large difference in binding energies between the neutron star and one of the two white dwarves makes this system very promising, but a direct measurement has yet to come. 
Alternatively, an interesting, yet indirect, test of the EP can be achieved via the $\gamma$ parameter 
(which enters the post-Newtonian expression $\eta=4\beta-\gamma-3$) 
by measuring differences in the Shapiro time delay between photons emitted from radio sources \cite{wei2015,nusser2016} or, 
more recently, between the first ever detected gravitational wave signal measured at different frequencies \cite{kahya2016}. 

\begin{figure}[!htbp]
\begin{center}
\includegraphics[width=.9\columnwidth]{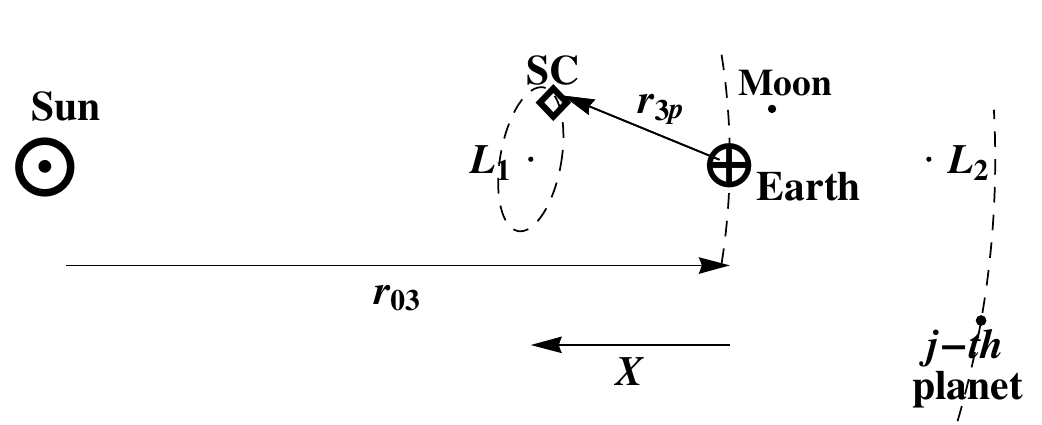}
\caption{Spacecraft ranging towards $L_1$ or $L_2$ as a means by which to test the SEP (not in scale). We calculate the SEP signature as a perturbation on the Earth's orbit around the the Sun (${\bf r}_{03}$) as well as on the spacecraft ranging (${\bf r}_{3p}$). We also include perturbations from other planets.}
\label{fig:diagram}
\end{center}
\end{figure}

The simplest form of EP violation for the body $i$ can be parametrised as follows \cite{damour1996,milani2002}
\begin{equation}\label{eq:general_ep_violation}
 m^\text{G}_i=m^\text{I}_i\, \left(1+\delta_i+\eta\,\Omega_i \right),
 \end{equation}
 where $m^\text{I}_i$ ($m^\text{G}_i$) is the inertial (gravitational) mass, 
$\delta_i\neq0$ is the WEP violation parameter, $\eta\neq0$ is the SEP violation parameter, also known as the \textit{Nordtvedt parameter}, and
 \begin{equation}
 \Omega_i=\frac{E_\text{g}}{m^\text{I}_i c^2}=-\frac{G}{2 m^\text{I}_i c^2}\iint \frac{  {\text{d}m^\text{G}_i}' {\text{d}m^\text{G}_i}''}
{\vert \mathbf r'-\mathbf r'' \vert},
\end{equation}
where $E_\text{g}$ is the self-gravity potential energy.
Therefore, $\Omega_i$ is the fraction of rest mass that contributes to self-gravity.
For instance, the Sun, Earth, and Moon have respectively 
$\Omega_0 = -3.52 \times 10^{-6}$, 
 $\Omega_\e= -4.64 \times 10^{-10}$, and $\Omega_\m = -1.88 \times 10^{-11}$ \cite{williams2009}.
A WEP violation prescribes that bodies with the same mass but different internal composition might fall at different rates [modelled by the parameter $
\delta_i$ in Eq.\;\eqref{eq:general_ep_violation}], while a SEP violation may be induced by differences in the bodies' self-gravity [collectively modelled by $
\eta\,\Omega_i$ in Eq.\;\eqref{eq:general_ep_violation}].

With experiments on ground, the typical $\Omega_i$ can be so small ($\lesssim 10^{-26}$) that only the WEP can effectively be tested. 
The only means by which the SEP can be constrained is evidently in space where the self-energies are much bigger.
The first measurement of the SEP's $\eta$ was proposed by Nordtvedt \cite{nordtvedt1968}. This experiment requires measuring the differential acceleration between the Earth and the Moon, both free falling in the Sun's gravity --the so-called \textit{Nordtvedt effect}. This differential acceleration is then
\begin{equation}
(\delta a_{\e\m}/a_\s)=\delta_\e-\delta_\m + \eta (\Omega_\e-\Omega_\m)\equiv\delta_{\e\m}+ \eta \Delta 
\Omega,
\end{equation}
where $a_\s$ is the mean acceleration of the Earth induces by the Sun's gravity. We may assume that the Earth and the Moon have different values of $\delta$ and $\Omega$. In principle, they may be characterised by different chemical compositions and different matter density distributions. Therefore, by accurately monitoring the Earth-Moon relative motion is possible to gain information about both the WEP and the SEP.
Over the last 45 years, the Lunar Laser Ranging (LLR) project has carried out a long sequence of measurements referred to as \textit{normal points} \footnote{The normal point is defined by the mean flight time of the photons received back on earth after being reflected by the Moon.}, with over 17,000 measurements in 2012 \cite{muller2012}.
With increasing precision on these measurements (from $\unit[20\text{--}30]{cm}$ in the 70's to currently $\approx\unit[1]{mm}$ \cite{murphy2012}), the final achieved root mean square (RMS)
uncertainty was $\sigma[\delta a_{\e\m}/a_\s) \approx 1.3 \times 10^{-13}$ \cite{williams2009}.
In order to subtract the WEP contribution, an experiment involving test masses with chemical composition similar to that of Earth and Moon yielded $\sigma[\delta a_\text{WEP}/a_\s)=1.4 \times 10^{-13}$  \cite{adelberger2009}. Combining these results, the best measurement of the RMS error associated to $\eta$ is currently $\sigma[\eta]=4.4 \times 10^{-4}$ \cite{turyshev2007, williams2009, adelberger2009}.

Alternative tests of the SEP were also proposed in the past. These experiments require the ranging between the Earth and another object orbiting around the Sun (not necessarily a 
planet). The main advantage is twofold: a longer baseline  ($\approx\unit[1]{AU}$ vs $\approx\unit[3\times10^{-3}]{AU}$) and 
 $\delta a/a_\s \propto \Omega_0$ instead of $\Delta \Omega$. 
This in turn implies a much bigger ranging signal amplitude (about three orders of magnitudes better than the 
Nordtvedt effect \cite{turyshev2004,milani2009}) and an increased precision on
$\eta$, since 
 $\Omega_0 \gg \Delta \Omega$.
One key mission is BepiColombo (BC) that will provide radio tracking data between the Mercury Planet Orbiter and the Earth \cite{milani2002,ashby2007}. 
The expected measurement precision on the SEP is $\sigma[\delta a/a_\s] \approx 10^{-11}$, which will be roughly 2 orders of 
magnitude worse than WEP measurements achieved by LLR and torsion balances experiments. However the parameter $\eta$ will be constrained with an accuracy of $10^{-5}\text{--}10^{-6}$ \cite{milani2002}, better than LLR. 
In fact, even if 
the time span and the precision of the data will be worse, 
 a bigger $\Omega_0$ and a stronger signal will certainly allows better measurements of $\eta$.
Testing of the SEP can also be investigated with the Earth-Mars \cite{anderson1996} or Earth-Phobos \cite{turyshev2004,turyshev2007,turyshev2010} ranging.

The Lagrangian points $L_4$ and $L_5$ were considered in a number of configurations (e.g.\, Sun-planet or
planet-satellite \cite{nordtvedt1968b,overduin2014,nordtvedt1970}). However, surprisingly enough, to our 
knowledge there has been no work done on the \textit{collinear} Sun-Earth Lagrangian points $L_1$ and $L_2$. 
This paper investigates the feasibility of using a radio tracking campaign towards one or more satellites in orbits around $L_1$ and $L_2$ to further constrain the SEP's $\eta$ parameter [see \fref{fig:diagram}]. 
The advantages of such a measurement will be discussed throughout the text. We anticipate that a measurement carried out for five years would be enough to reach the LLR constraint. 
The analysis of ranging data from current and future missions would also be able to further improve the constraint.

The structure of this paper is as follows. 
After some preliminaries (\sref{sec:strawman} and \sref{sec:notation}), 
we calculate the perturbation of the Earth's heliocentric trajectory due to SEP violation
(\sref{sec:EMB}), and then we work out the signature of the SEP violation on the relative Earth-spacecraft distance for a spacecraft placed around $L_1$ or $L_2$ 
(\sref{sec:lagrange}).
In \sref{sec:prediction} we do numerical simulations to forecast the figure-of-merit for the proposed $\eta$-measurement towards $L_1$ or $L_2$ and compare this with the most recent LLR measurement and the expected performance of BC.


\section{Measuring $\eta$ by spacecraft ranging towards the nearby Lagrangian points}\label{sec:math}

\subsection{Straw man calculation}\label{sec:strawman}

Let us make a preliminary estimation of the expected RMS error of $\eta$ for a spacecraft (SC) in orbit around the Sun-Earth $L_1$/$L_2$ points.
As masses and self-gravity energies of planets and SCs are negligible with respect to the Sun, by all means there is 
no difference between the dynamics of a SC and that of a planet. 
Therefore, the SEP signature is always proportional to $\Omega_0$, the self-energy of the Sun.
The range baseline, on the contrary, is quite small compared to a planetary mission where typical range distances are $\approx\unit[1]{AU}$ or more. The $L_1$/$L_2$ points are at about $\unit[0.01]{AU}$ from the 
Earth (inward and outward, respectively), therefore the amplitude of the signal should be around the 
same order of the Nordtvedt effect, which is $\approx\unit[13]{m}$ over the Earth-Moon distance. 
In fact, as the expected signal for a planetary mission is $\unit[10^2\text{--}10^3]{m}$, 
with a factor $\approx100$ shorter baseline we foresee a signal of $\approx\unit[1\text{--}10]{m}$.
The expected precision on differential acceleration for a SEP test around $L_1$/$L_2$ is therefore $\sigma[\delta a/a_\s]\approx100\times 10^{-11}= 
10^{-9}$, which gives,
by dividing by $\Omega_0$, the final result of $\sigma[\eta]=3.4\times 10^{-4}$ -- incidentally the same order of magnitude of LLR.
We therefore deduce that a radio tracking campaign towards a SC orbiting around $L_1$ or $L_2$, 
despite its smaller baseline, could be a valid measurement setup for testing the SEP, 
which is both alternative to LLR and complementary to planetary missions.


\subsection{Notation and reference frames}\label{sec:notation}

Before going through the actual calculation of the expected SEP violation signal, it is worth reviewing the notation, and defining the relevant reference frame. 
Hereafter, 
the index $j=1,\dots,8$ identifies the $j$th-planet of the solar system, $j=0$ being the Sun, $j=3$ the Earth-Moon 
system, and so on. We also assume that the Earth's position coincides with the Earth-Moon barycentre as the motion of the Earth about it is a very small contribution to our signal: this effect can be affectively neglected without affecting our results ($m_\e \approx 81\, m_\m$).  
We denote $\Omega_3=\Omega_\e+\Omega_\m$ and $m_3=m_\e+m_\m$.

We will work in the heliocentric reference frame, where 
the unit vectors for the $j$th-body are: $\ve u_r^j$  for the radial, $\ve 
u_t^j$ for the along-track, and $\ve u_z^j$ for the out-of-plane components. 
We denote the position of the $j$-th body in a certain coordinate system with $\ve r_j$, and define $\ve r_{ij}=\ve r_j-\ve r_i$ with $r_{ij}=\vert \vert \ve r_{ij} \vert \vert$.
We assume circular and co-planar orbits for all bodies.
In addition, the orbital frequencies of the planets are defined as follows ($\mu_j=G\, m_j^\text{G}$)
\begin{equation}
n_j =\sqrt{\frac{\mu_0+\mu_j}{r_{0j}^3}},
\end{equation}
where $r_{0j}$ is the semi-major axis of the $j$-th planet.
We also introduce the orbital phase as a function of time $\Phi_j(t)=n_j t+ \varphi_j$, where $\varphi_j$ is the phase angle at $t=0$.
Finally, for each pair of planets $i$ and $j$ we define the difference of their orbital frequencies, $n_{ji} = n_i - n_j$, and the difference of their orbital phase, $\Phi_{ji}(t)=\Phi_i(t)-\Phi_j(t)$.


\subsection{SEP signature on the dynamics of the Earth around the Sun}\label{sec:EMB}

Let us begin with calculating the induced SEP effect on the dynamics of the Earth in orbit around the Sun. 
The equation of motion of the Earth in the heliocentric frame, including all planetary perturbations, is given by \cite{milani2002}
\begin{equation}\label{eq:bar1}
\begin{split}
\ddot {\ve r}_{03} &= - \dfrac{\mu^\star}{r_{03}^2}\ve u_r^3+\sum_{j \neq 0,3} \mu_j \left(\dfrac{\ve 
r_{3j}}{r_{3j}^3}-  \dfrac{\ve r_{0j}}{r_{0j}^3} \right) \\
 &\quad+ \eta\, \sum_{j \neq 0,3} \mu_j \left(\Omega_3 \dfrac{\ve r_{3j}}{r_{3j}^3}-\Omega_0  \dfrac{\ve 
r_{0j}}{r_{0j}^3}\right),
\end{split}
\end{equation}
where $\mu^\star= \mu_0+\mu_3+\eta(\mu_3 \Omega_0+\mu_0 \Omega_3)$. 
In the above equation of motion,
the first sum, which is a planetary tidal contribution, 
does not depend on $\eta$ at first order. However, as the planets' trajectories and masses 
are affected by measurement uncertainty, this term turns out to be crucial for
parameters estimation, and in particular for this particular ranging measurement.
In this work we will neglect any planetary term as it is second order, and also any planetary uncertainty propagated onto our SEP forecast --
we reserve to quantify this in the future.

As $\Omega_3 \ll \Omega_0$, we also neglect the term $\propto \Omega_3$ in the second sum.
We seek a solution -- the heliocentric position of the Earth as a function of $\eta$ and time -- for the above equation of motion in the form
\begin{equation}\label{eq:r03}
\ve r_{03}= (R+ \eta \,\delta x_3) \ve u_r^3 + \eta \,\delta y_3\, \ve u_t^3 + \mathcal{O}(\eta^2),
\end{equation}
where $R=\unit[1]{AU}$ (we neglect the orbital eccentricity) and $\delta x_3$ and $\delta y_3$ are evidently the radial and along-track components of the orbital perturbation due to the SEP violation.
Linearising \eref{eq:bar1} for small perturbations gives the following system of Hill-Clohessy-Wiltshire  \cite{cw1960} perturbed equations
\begin{subequations}\label{eq:emb}
\begin{align}
 \delta \ddot x_3 -2  n_3 \, \delta \dot y_3- 3  n_3 ^2 \, \delta x_3 &= -\Omega_0 \sum_{j \neq 0,3} 
\dfrac{\mu_j}{r_{0j}^2} \cos \Phi_{j3}, \\
  \delta \ddot y_3 +2  n_3 \, \delta \dot x_3 &= \Omega_0 \sum_{j \neq 0,3} \dfrac{\mu_j}{r_{0j}^2} 
\sin \Phi_{j3}.
\end{align}
\end{subequations}
Solutions of the above equations can be cast in the following general form
\begin{equation}\label{eq:solg}
\delta \ve r_3 =  \delta \ve {\hat r}_3+\Omega_0 \sum_{j \neq 0,3}\frac{\mu_j}{r_{0j}^2} \, \left[ {\cal 
R}_{j3} \cos \Phi_{j3}\, \ve u_r^3 +  {\cal T}_{j3} \sin  \Phi_{j3} \,\ve u_t^3\right].
\end{equation}
In other words, the solution is the sum of a homogeneous solution, $\delta \ve {\hat r}_3=(\delta \hat x_3, \delta \hat y_3)$ 
\footnote{
The homogeneous solution for the perturbation to the Earth's heliocentric dynamics 
can be written as follows $\delta \hat x_3 =A \cos n_3 t+B \sin n_3 t + C$, and
$\delta \hat y_3 = 2 B \cos n_3 t -2 A \sin n_3 t- 3/2 C \,n_3 t+D$,
where the coefficients $A,B,C,D$  can be determined by imposing the heliocentric initial conditions on the Earth's position and velocity.} 
plus an inhomogeneous solution that is expressed as a series of sine/cosine functions that depend on the gravitational interaction with the other planets. 
Their amplitudes are $\Omega_0 \mu_j / r_{0j}^2\times \{{\cal R}_{j3}, {\cal T}_{j3}\}$, where the coefficients
\begin{subequations}
\begin{align}\label{eq:RT_coeffs}
{\cal R}_{j3} &=  \dfrac{1+2 n_3/n_{j3}}{n_{j3}^2-n_3^2}, \\
{\cal T}_{j3} &= -\dfrac{1+ 2 n_3/ 
n_{j3} + 3 n_3^2/n_{j3}^2}{n_{j3}^2 - n_3^2},
\end{align}
\end{subequations}
depend only on the Earth's orbital frequency and its difference with the planets' orbital frequencies.
Numerical values for all these amplitudes, $\Omega_0 \mu_j / r_{0j}^2\times \{{\cal R}
_{j3}, {\cal T}_{j3}\}$, which once multiplied by $\eta$ give the observable SEP signal in the Earth's dynamics, are reported in \tref{tab:perts_coefficients}.


\subsection{SEP signature on the spacecraft ranging}\label{sec:lagrange}

We now go through the calculation of the signal due to SEP violation in the SC ranging. 
We place a spacecraft on a Lissajous orbit around 
one of the two nearby Lagrangian points of the Sun-Earth system and
we calculate the perturbation on the relative motion between Earth and the SC.
 
The position, $X$, of a collinear Lagrangian point ($L_1$, $L_2$, or $L_3$) is given by the equilibrium between the real gravitational forces of the Sun and Earth, and the inertial forces (essentially, a centrifugal force). In the Earth's reference frame, this equilibrium is given by the following equation
\begin{equation}\label{eq:L1}
-\frac{\mu_0}{ \vert R-X \vert^3} (R - X) + \mu_3 \left(\frac{X}{ \vert X \vert ^3} -\frac{1}
{R^2}\right)+  n_3 ^2 (R-X) =0,
\end{equation}
which has three solutions: $X_{1,2}\approx\pm\unit[0.01]{AU}$ that correspond to $L_1$ and $L_2$, 
and $X_3\approx\unit[2]{AU}$ that corresponds to $L_3$. 
We will consider only the case of $L_1$ and $L_2$ as these are the spots where many missions fly to.

Consider a SC, hereafter identified with the index $p$,  near $L_1$ (or $L_2$). Its mass and self-gravity 
energy are negligible with respect to those of the Sun and all planets.
We are interested in deriving the trajectory of the SC relative to Earth and see how this is affected by a SEP violation at first order. 
First, we write the SC's equation of motion
relative to the Sun [see \eref{eq:bar1} where we substitute $(\Omega_3,\mu_3,\ve r_{03},\ve r_{3j}) \rightarrow (0,0,\ve r_{0p},\ve r_{pj})$],
and then we subtract it from the Earth's equation of motion to finally derive the 
relative motion, $\ve r_{3p}$, between the SC and Earth, which is given by
\begin{equation}\label{eq:L1_rel}
\begin{split}
\ddot {\ve r}_{3p} &= - \mu_0 \left( \dfrac{\ve r_{0p}}{r_{0p}^3}-  \dfrac{\ve r_{03}}{r_{03}^3}\right) 
-\mu_3 \dfrac{\ve r_{3p}}{r_{3p}^3}+ \\
&\quad+ \sum_{j\neq 0,3} \mu_j \left(\dfrac{\ve r_{pj}}{r_{pj}^3}-\dfrac{\ve r_{3j}}{r_{3j}^3}\right)+ \eta\, 
\Omega_3 \sum_{j \neq 3} \mu_j \dfrac{\ve r_{j3}}{r_{j3}^3},
\end{split}
\end{equation}
where $\ve r_{0p}=\ve r_{03}+\ve r_{3p}$. It is worth noting that we are solving the equation of motion for the observed SC ranging, $\ve r_{3p}$. 
In turn, this will depend explicitly on $\eta$ through the last sum in the equation, 
but also implicitly through the relative distance between Earth and Sun, $\ve r_{03}$, which is given by \eref{eq:r03}.
It is this implicit term that will dominate the forecast of the SEP measurement, not the explicit one, which is proportional to $\Omega_3\ll\Omega_0$.
Much like before, the first sum represents the tidal interaction with the other planets, which we neglect in this case.
We introduce the following constants
\begin{subequations}
\begin{align}
n _z &= \sqrt{ \frac{\mu_0}{(R-X)^3} + \frac{\mu_3}{\vert X \vert ^3}}, \\
Q &=\frac{\mu_0}{(R-X)^3}-\frac{\mu_0}{R^3}.
\end{align}
\end{subequations}
Analogously to what we have done before, we seek for a solution as follows 
\begin{equation}
\ve r_{3p}=(-X+\eta \,\delta x)\ve u_r^3+\eta \,\delta y \ve u_t^3+\mathcal{O}(\eta^2),
\end{equation}
which yields, with a bit of mathematics [Eqs \eqref{eq:solg}, \eqref{eq:L1}, and \eqref{eq:L1_rel}], the equations for the radial and along-track SC positions relative to Earth
\begin{subequations}\label{eq:rel}
\begin{align}
\begin{split}
\delta \ddot x -2  n_3  \delta \dot y -(n_3^2+2 n_z^2) \delta x  &=  2 \, Q \delta x_3 +  {\cal D}_r
\end{split}\\
\begin{split}
\delta \ddot y +2  n_3  \delta \dot x -(n_3^2- n_z ^2) \delta y  &= -Q \delta y_3+ {\cal D}_t,
\end{split}
\end{align}
\end{subequations}
where  the \textit{direct} terms are given by
\begin{subequations}
\begin{align}
{\cal D}_r &=  \Omega_3\,\dfrac{\mu_0}{R^2}+\Omega_3\sum_{j\neq 0,3} \mu_j \dfrac{R- r_{0j} \cos \Phi_{j3}}{r_{j3}^3},\\
{\cal D}_t & =  \Omega_3 \sum_{j\neq 0,3} \mu_j \dfrac{ r_{0j} \sin \Phi_{j3} }{r_{j3}^3},
\end{align}
\end{subequations}
which, as said earlier, are evidently small as $\Omega_3\ll\Omega_0$.
It is worth mentioning that, among all, the solar term is the biggest one, 
but it leads to an unobservable radial ``permanent tide" with a DC amplitude of $\approx 8 \eta$.
From \eref{eq:rel}, we deduce that the motion of the SC relative to Earth consists of a set of perturbed Lissajous orbits (see e.g.\ Ref.\ \cite{richardson1980}),
the perturbation being indirectly generated by the  displacement $\delta \ve r_3$ of \eref{eq:solg} in the Earth's position due to the SEP violation.

We now search for a solution of \eref{eq:rel} in the form 
\begin{subequations}\label{eq:solrel}
\begin{align}
\delta x &= \delta  x'+ \delta \hat x+ \Omega_0 \sum_{j \neq 0,3}  a_x^j \cos \Phi_{j3},\\
\delta y &= \delta y' + \delta \hat y+  \Omega_0 \sum_{j \neq 0,3}  b_y^j \sin \Phi_{j3}, 
\end{align}
\end{subequations}
where 
$\delta  x'$ and $\delta  y'$ are the homogenous solutions of  \eref{eq:rel}, 
which depend on the initial position $\ve r_{3p}(0),$ and velocity $\dot {\ve r}_{3p}(0)$ of the SC with respect to Earth. Instead, $\delta 
\hat x$ and $\delta \hat y$ are the homogeneous solutions, $\delta \ve {\hat r}_3$, for the perturbation of the Earth's orbit in \eref{eq:solg}. 
Analytical details about the solution of \eref{eq:rel} are reported in  Appendix \ref{app:sc_eom}.

The numerical values for the coefficients $\Omega_0 \times \{a_x^j,b_y^j\}$ of the inhomogeneous solution are reported 
in \tref{tab:perts_coefficients} and represent the orbital perturbations due to the other planets. Therefore, as a SEP violation affects directly the Earth dynamics and this is included in the SC ranging, in turns the SC-Earth relative distance is a physical observable for a SEP violation, even though the SC itself is by all means considered as a test mass (point-like source with no self-energy).

\begin{figure}[!htbp]
\begin{center}
\includegraphics[width=\columnwidth]{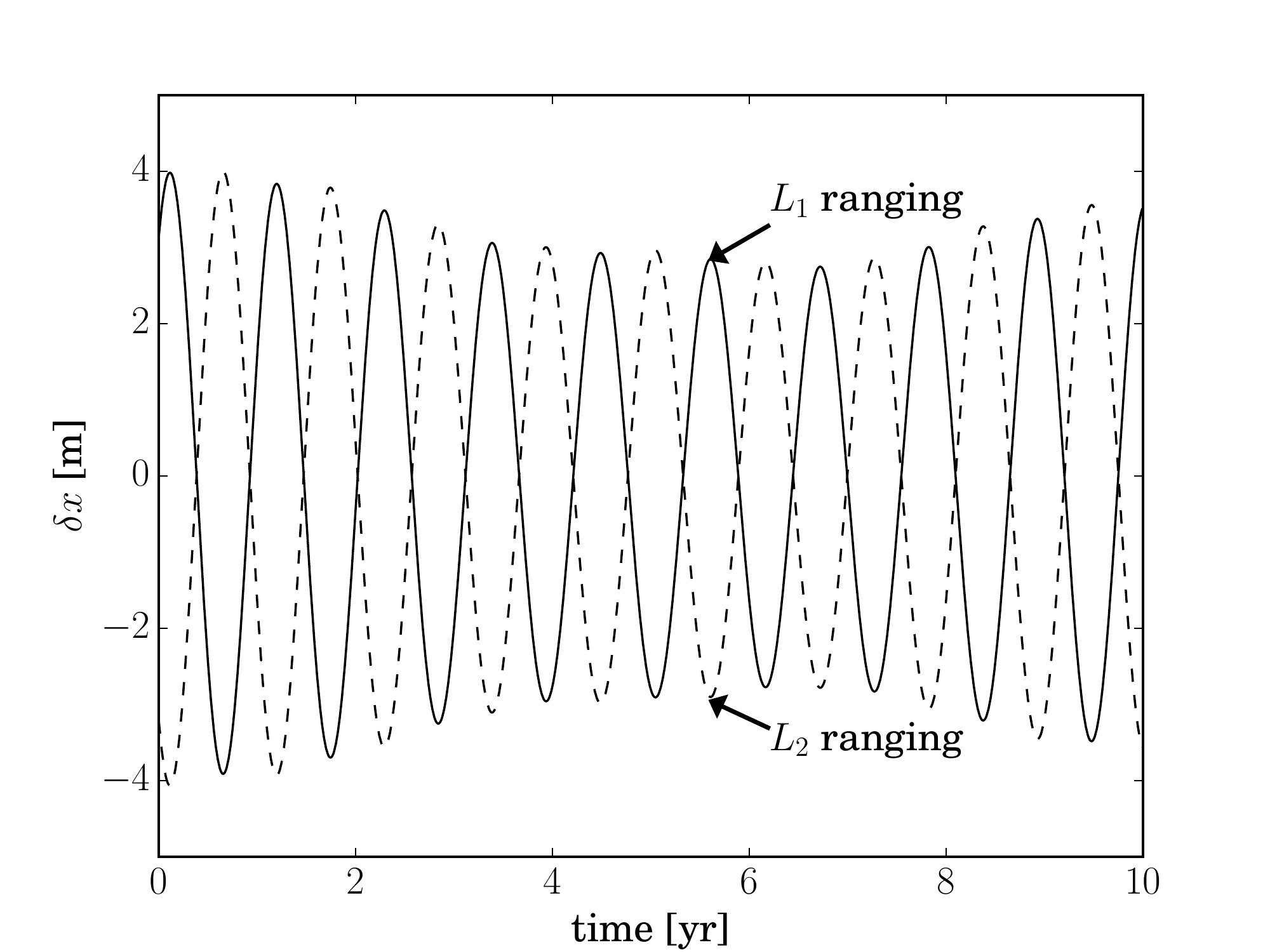}
\caption{Expected perturbation to the range signal towards a spacecraft in orbit around $L_1$ (solid line) and $L_2$ (dashed line) due to the SEP violation. The two signals have opposite sign.}
\label{fig:ranging}
\end{center}
\end{figure}

It is also worth mentioning that as $L_1$ and $L_2$ are placed approximately at the same distance from Earth and 
$\vert X \vert \ll r_{03}$, the factor $Q$ in Eq.\ \eqref{eq:rel} can be approximated as follows
\begin{equation}
Q \approx 3   \frac{\mu_0}{R^4} X.
\end{equation}
Therefore we get the interesting result that the SEP signature on the range signals towards $L_1$ and $L_2$ are quasi-identical in shape, but with 
opposite sign (since $X$ changes sign between $L_1$ and $L_2$).
In \fref{fig:ranging} we show an example of a SEP perturbation on the ranging towards $L_1$ and $L_2$.

\section{Experimental forecast}\label{sec:prediction}

In the previous section we set up our mathematical framework, 
we are now ready to forecast the figure-of-merit for a ranging experiment towards $L_1$ or $L_2$. 
The SEP signature is contained in the perturbation of the Sun-Earth distance, $\delta x_3$, and 
in the perturbation of the Earth-SC ranging, $\delta x$.
We calculated these two quantities in the previous section, specifically with Eqs \eqref{eq:solg} and \eqref{eq:solrel}.
We report all the planetary contributions to both $\delta x_3$ and $\delta x$ in \tref{tab:perts_coefficients}.
A plot of $\delta x$ is shown in \fref{fig:ranging}, where it is clear that the effects of $L_1$ and $L_2$ have opposite signs. 
The largest amplitude is due to Jupiter and corresponds to $\Omega_0 \,a_x^5 \approx\unit[3.7]{m}$. 
By contrast, the Nordtvedt effect on the lunar orbit gives $\approx\unit[13]{m}$ \cite{damour1996} and its frequency is about 10 times larger than the frequencies that are typically involved in our measurement setup.

\begin{table}[!htbp]
\begin{ruledtabular}
\begin{tabular}{llcccc}
Planet  & Period & \multicolumn{2}{c}{Sun-Earth}  &  \multicolumn{2}{c}{Earth-SC ($L_1$)} \\
\hline
& $2 \pi/ \vert n_{j3} \vert$  & \multicolumn{2}{c}{$\Omega_0 \mu_j / r_{0j}^2 \times$}  &  
\multicolumn{2}{c}{$\Omega_0 \times$} \\
& & ${\cal R}_{j3}$ & ${\cal T}_{j3}$ & $a_x^j$ & $b_y^j$ \\
& [d]  & [m] & [m] & [m] & [m] \\
Mercury  &  115.9 & -0.0239 & 0.0436 & 0.0002 &  -0.0004 \\ 
Venus     &  582.9 &-8.8829 & -22.0822  & 0.0850 & -0.2126 \\ 
Mars       &  747.3 & 0.4649 & -1.6115 & -0.0047 & 0.0158 \\ 
Jupiter   &  398.8 &366.257& -777.6860  & {\bf -3.6544} &7.6681 \\ 
Saturn   &  378.1 &76.0374 & -155.6470 & -0.7582 & 1.5439 \\ 
Uranus   &  369.7 &7.9818 &-16.0921  &-0.0796 & 0.1601 \\ 
Neptune &  367.5 &7.4410 &-14.9426 &  -0.07419 & 0.1488 \\ 
\end{tabular}
\end{ruledtabular}
\caption{Planetary contributions to the perturbation of a SC's orbit around $L_1$, due to a SEP violation. We report the typical period (col.\ 2), 
the radial and along-track signals for the perturbation of the Sun-Earth distance [see Sec.\ \ref{sec:EMB}] (cols 3 and 4), 
and the radial and along-track signals for the perturbation of the Earth-SC ranging [see Sec.\ \ref{sec:lagrange}] (cols 5 and 6). 
Jupiter contributes to much of the radial signal with $\approx\unit[3.7]{m}$. 
Note that the actual observable is the
series of the coefficients $\Omega_0 a_x^j\cos{\Phi_{j3}}$ along the radial component, where the measurement error is small enough to test the SEP. 
Along-track components have much bigger measurement errors.}
\label{tab:perts_coefficients}
\end{table}

We refer the reader to Appendix \ref{app:sc_eom} for the SC's equation of motion relative to Earth, and its general solutions. 
According to the unstable dynamical behaviour of the collinear Lagrangian points, 
the complete homogeneous solutions of \eref{eq:rel} are divergent. In practice, this is compensated by correcting the SC's orbit from time to time by pushing it back roughly along the radial position. 
Since accurate modelling of feedback-controlled orbital dynamics is far beyond the scopes of this paper, we decided to avoid drifts by imposing that all coefficients of real exponentials should be zero 
[the constraint $A_1,A_2 \equiv 0$ in \eref{eq:semistable}].
Consequently, there are two initial conditions that are linearly dependent on the other two. 
In calculating the expected perturbation to the SC ranging, we keep free the following set of parameters: ({\textit{i}) the SEP's $\eta$ parameter,
({\textit{ii}) the initial position and velocity of the Earth relative to the Sun, $\ve r_3 (0), \ve{\dot r}_3 (0)$, and 
(\textit{iii}) the initial position of the SC relative to Earth, $\ve r_{3p} (0)$. As each of the two relative motions has two degrees of freedom, this makes 1+6 parameters in total, and we collect those parameters in the following vector $\vec{a}=\{\eta,\vec{\theta}\}$, 
where the $\eta$ parameter is the focus of our analysis and $\vec{\theta}=\{\ve r_3 (0), \ve{\dot r}_3 (0), \ve r_{3p} (0)\}$ 
is the set of all initial conditions that are evidently a nuisance for our analysis.
The model of the perturbation of the ranging towards $L_1$ can therefore be written as the analytical function
$\rho\equiv\eta\,\delta x(t, \vec{a})$.

We can now calculate the expected RMS error on $\eta$, marginalised over the nuisance.
We assume we have $N$ equally-spaced observations of the SC's range distance, over a total observation of $T=\unit[5]{yr}$, sampling interval $\delta t=T/N=\unit[1]{h}$,
and range error $\sigma_i$ ($\unit[1]{h}$ integration time). Therefore, the expected Fisher information matrix, the so-called \textit{normal matrix}, is by definition
\begin{equation}
F_{jk} = \sum_{i=0}^{N-1}  \frac{1}{\sigma_i^2} \frac{\partial \rho (t_i,\vec{a})}{\partial a_j} \frac{\partial \rho 
(t_i,\vec{a})}{\partial a_k}+\sigma^{-2}[a_j] \delta_{jk}.
\end{equation}
This has to be evaluated at some fiducial values $\vec{a}_0$ 
-- we assume $\eta_0=0$ (SEP is valid), initial position and velocity of the Earth at a given epoch 
\footnote{Any change in the starting epoch induces a corresponding change of phase in the perturbation of the range signal.},
and some arbitrary initial position for the SC.
Whenever available, we apply Gaussian priors independently on each of the initial conditions, $\sigma^2[a_j]$, and include these in the Fisher matrix. The marginalised error on $\eta$ is therefore given by $\sigma[\eta]=\sqrt{(F^{-1})_{00}}$. 

In order to make our forecast, we distinguish between two possible scenarios.
In the \textit{realistic scenario} (A) we use a nominal range error typical for two-way ranging in the $X$-band, $\sigma_i=\unit[0.1]{m}$ ($\unit[1]{h}$ integration time)
\footnote{As obtained from the $K_a$-band range error 
$\sigma_i=0.15 \sqrt{300/\delta t}\approx\unit[0.04]{m}$ ($\delta t = \unit[1]{h}$) \cite{iess2001,schettino2015,cicalo2016}, 
degraded by a conservative factor of 2.5, 
owing to the lower frequencies typical of the $X$ band.}.
Additionally, we assume the following prior uncertainties on the orbital initial conditions: 
(\textit{i}) $\unit[2]{m}$ and $\unit[3\times10^{-5}]{m/s}$ for the Earth's heliocentric radial position and velocity, 
from a great abundance of radio tracking data 
\cite{kaplan2015}; 
(\textit{ii}) $\unit[145]{m}$ for the Earth's heliocentric along-track position as this is less well constrained \cite{kaplan2015}; 
(\textit{iii}) no prior both on the Earth's heliocentric along-track velocity as this is very weakly constrained by current data, and on the parameters of the SC's orbit relative to Earth. 
In the \textit{optimistic scenario} (B) we use the range error
typical of the $K_a$ band, $\sigma_i=\unit[0.04]{m}$ ($\unit[1]{h}$ integration time), 
as well as a factor 10 improvement in the knowledge of the Earth's initial position and velocity, $\unit[0.2]{m}$ and $\unit[3\times10^{-6}]{m/s}$, 
which is likely to be achieved in the near future.

\begin{table*}[!htbp]
\begin{ruledtabular}
\begin{tabular}{lllllll}
Experiment & Range baseline [AU] & Range error [m] & Time span [y] & $\sigma[\eta]/10^{-4}$ & Note & Ref.\ \\
\hline
$L_1$               & 0.01 & 0.1$^\text{A}$, 0.04$^\text{B}$ & 5 & 6.4$^\text{A}$, 2.0$^\text{B}$ & forecast & this work \\
$L_2$               & 0.01 & 0.1$^\text{A}$, 0.04$^\text{B}$ & 5 & 7.0$^\text{A}$, 2.1$^\text{B}$ & forecast & this work \\
$L_1$ + $L_2$ & 0.01 & 0.1$^\text{A}$, 0.04$^\text{B}$ & 5 & 4.8$^\text{A}$, 1.7$^\text{B}$ & forecast & this work \\
LLR                                & $2.6\times10^{-3}$ & 0.2-0.001 & 46 & 4.4 & current best measured  & \cite{turyshev2007,williams2009,adelberger2009}\\
BepiColombo                 & 0.6-1.4 & 0.24 & 1 & $< 0.1$ & expected upper limit & \cite{tommei2015, schettino2015}
\end{tabular}
\end{ruledtabular}
\caption{SEP testing performances for the ranging towards $L_1$/$L_2$, compared with LLR and BC. 
Our forecast figure-of-merit is the uncertainty on the SEP's parameter, $\eta$, 
for $L_1$ alone, $L_2$ alone, and
$L_1$ and $L_2$ combined (this work);
LLR (current best measured); BC (expected upper limit).
For the ranging towards $L_1$/$L_2$, we assume a realistic scenario (A) with current range error capabilities ($\unit[0.1]{m}$) and current knowledge of the Earth's initial radial position and velocity ($\unit[2]{m}$ and $\unit[3\times10^{-5}]{m/s}$), and an optimistic scenario (B) with improved range capabilities ($\unit[0.04]{m}$) and a factor 10 improvement in the knowledge of the Earth's initial radial position and velocity.
The ranging towards $L_1$/$L_2$ would allow us to reach   
the performances of LLR in both scenarios.
We mention possibilities for further improvement in the final discussion.}
\label{tab:FOM}
\end{table*}

Our predicted figure-of-merit in both measurement scenarios is reported in \tref{tab:FOM}, 
where we compare these figures with the current best measurement from LLR and the expected performance of BC. 
In the realistic scenario and integrated for 5 years,
we forecast $\sigma[\eta]=6.4\times 10^{-4}$ for a single SC around $L_1$ and $4.8\times 10^{-4}$ for a combined measurement of two SCs around $L_1$ and $L_2$. This is just above $4.4\times 10^{-4}$ achieved by LLR measurements over more than 40 years. 
In the optimistic scenario, the forecast yields $2.0\times 10^{-4}$ and $1.7\times 10^{-4}$ respectively for $L_1$ and $L_1$+$L_2$, again integrated over 5 years. 
It is also worth mentioning that a time span of one year would already be enough to get $\approx3\times 10^{-4}$. 
The expected performance of BC is of course at least an order of magnitude better \cite{milani2002, tommei2015, schettino2015}, 
but we do envisage here the difficulties related to such a measurement as compared to a relatively simple measurement towards the collinear Lagrangian point 
and a fairly easy integration of the signal over time thanks to the many SCs that could possibly fly around $L_1$ and $L_2$.

In doing this exercise, we identified two major sources of performance degradation.
 The first one is the range error that mostly depends on the frequency band of the SC transponder used for the modulation and integration of the Doppler signal. 
As $K_a$ frequencies are typically 2-3 times larger than in the $X$ band and the range error scales inversely with frequency, 
we get a similar improvement factor in the range error. 
It is worth noting that a number of satellites are now adopting $K_a$ for their tracking. 
The second source of degradation is the knowledge of the Earth's ephemerides. 
These are determined through spacecraft tracking of the the many missions in the solar system and through observation of reference astrophysical sources (e.g.\ quasars),
therefore the Earth's ephemerides are better and better constrained over time. 
We realised that the Earth's position, compared to velocity, had the dominant effect on the figure-of-merit -- the effect of velocity was indeed negligible.


\section{Discussion}\label{sec:conclusions}

We investigated the feasibility of a radio tracking campaign towards the two nearby Lagrangian points ($L_1$ or $L_2$) to test the SEP. 
Our figure-of-merit is the measurement uncertainty on the SEP parameter, $\eta$, that serves as the predicted 1-$\sigma$ upper limit on the SEP. 
We assumed a nominal measurement of five years, with cadence of one sample per hour, and nominal range error of $\unit[0.10]{m}$ or $\unit[0.04]{m}$ depending on the range precision. 
In our forecast analysis we included also the initial conditions of the Earth's orbit and the SC's orbit, we applied some prior knowledge of their values coming from independent measurements (essentially the Earth's radial position and velocity), and marginalised over these.
The expected marginalised uncertainty on $\eta$, via ranging towards $L_1$, 
gives $\sigma[\eta]=6.4\,(2.0)\times10^{-4}$ (5 years integration time),
in a realistic (optimistic) scenario, 
but it improves to $\sigma[\eta]=4.8\,(1.7)\times10^{-4}$ for a combined measurement towards $L_1$ and $L_2$. In the optimistic scenario,
 a single measurement of one year would already be enough to reach $\approx3\times10^{-4}$. All these figures are comparable with LLR, and just an
  order of magnitude below the expected performance of the future mission towards Mercury, BC. 
However, the limits of our forecast boil down to the current knowledge of the Earth radial position and the SC range error that determine our realistic and optimistic scenarios.
Moreover, in this work we did not consider a possible degradation of our forecast
owing to uncertainties in planetary masses and ephemerides. 
These errors might introduce spurious signals that would correlate with the SEP signal we are looking for. 
However, given the small baseline ($\unit[0.01]{AU}$) as compared to distances between planets, these signals are expected to be very small.
A detailed calculation to include these effects will be done in future work.

We point out that there are some key experimental advantages of $L_1$/$L_2$ over other experiments. We list them as follows. 
(\textit{i}) From the dynamical point of view, the SC's orbit would appear from Earth quasi-static in both the radial and along-track components.  
(\textit{ii}) The SC is by all means a test mass with no self-gravity, no figure effects are present and the dynamical modelling is much easier.
(\textit{iii}) From the point of view of radio tracking, the SC would be always visible from Earth and the measurement range would again be in more control, again helping a lot with the systematics. 
(\textit{iv}) As there is no potential limit to the experiment duration $T$ as long as the SC is kept in a stable orbital configuration around the Lagrangian point, 
the SEP signal will integrate as $\propto 1/\sqrt{T}$. 
(\textit{v}) With a number of missions flying around the Lagrangian points, information from different SCs, 
even at different epochs, can be combined and the performances will scale as the inverse square root of the number of experiments involved. 
(\textit{vi}) The radio tracking technology keeps improving with time and it is very likely that the range error will improve by at least an order of magnitude in the future. 
(\textit{vii}) Missions towards the Lagrangian points are generally cheaper than interplanetary ones.

Finally, we do not advocate a dedicated experiment to test the SEP, rather we do suggest using current data and equipping future missions with radio transponders that are accurate enough for the purpose of testing the SEP.
One critical aspect of such a measurement might be the ability to compensate for the radiation pressure from the Sun that would otherwise perturb the SC orbit
and therefore degrade the SEP measurement. 
Employing an on-board accelerometer would definitely benefit the subtraction of this unwanted noise source. 
At the time of writing this paper, a mission that would match this requirement is LISA Pathfinder, currently in science operations around $L_1$.
As a concluding remark, the ranging towards $L_1$/$L_2$ would serve as a direct test of the SEP, potentially less prone to systematic errors and independent from other experiments, and at least comparable in terms of performances achieved in a relatively short time span.

\begin{acknowledgments}
GC acknowledges support from the Beecroft Institute for Particle Astrophysics and Cosmology, and 
Oxford Martin School.
FDM acknowledges the advice and support of A.\ Milani and G.\ Tommei (Department of Mathematics, University of Pisa) during work on the Mercury Orbiter Radio-science Experiment on board BepiColombo, and N.\ Ashby and P.\ Bender (University of Colorado, Boulder) for fruitful interaction on the development of analytical models.
GC thanks D.\ Alonso and S.\ Naess (Department of Physics, University of Oxford) for a useful discussion on Fisher forecasting and accounting for systematics errors.
The authors thank E.\ Pitjeva (Institute of Applied Astronomy, Russian Academy of Sciences, St.\ Petersburg) and L.\ Imperi (Dip. di Ing. Meccanica e Aerospaziale,
Universit\`{a} degli Studi di Roma ``La Sapienza'') for discussions and clarifications with regarding the prior knowledge of the Earth's ephemerides.
The authors finally thank the anonymous referee for their valuable review, which provided improvements and suggestions for this paper and future work.
\end{acknowledgments}


\appendix

\section{Solution for the spacecraft trajectory relative to Earth}\label{app:sc_eom}
We consider the system of equations
\begin{align}
\delta \ddot x -2  n_3  \delta \dot y -(n_3^2+2 n_z^2) \delta x & =  f_r, \\
\delta \ddot y +2  n_3  \delta \dot x -( n _3^2- n_z ^2) \delta y  &=  f_t,
\end{align}
which admit the following homogeneous solutions \cite{richardson1980} 
\begin{align}\label{eq:semistable}
\delta \hat x &= A_1 e^{\lambda t}+ A_2 e^{-\lambda t}+ A_3  \cos  n _{xy}t + A_4 \sin  n _{xy}t, \\
\delta \hat y &= q\,A_1 e^{\lambda t}- q\, A_2 e^{-\lambda t}+ k\,A_3  \sin  n _{xy}t - k\,A_4 \cos  n _{xy}t,
\end{align}
with 
\begin{align}
n _{xy}&=\sqrt{\dfrac{2  n_3 ^2- n _z^2+\sqrt{9  n _z^4-8  n_3 ^2  n _z ^2}}{2}}, \\
\lambda &=\sqrt{\dfrac{-2  n_3 ^2+ n _z^2+\sqrt{9  n _z^4-8  n_3 ^2  n _z ^2}}{2}},  \\
q&= \dfrac{\lambda^2-n_3^2-2 n_z^2}{2 \lambda n_3}, \qquad k=-\dfrac{n_{xy}^2+n_3^2+2 n_z^2}
{2 n_{xy} n_3}.
\end{align}
The coefficients $A_1,...,A_4$ depend, of course, on the initial conditions. The exponential terms in \eref{eq:semistable} imply 
that in general orbits are not closed and therefore they become unstable. Homogeneous solutions can be forced to be stable with a particular choice of 
initial conditions that produce $A_1=A_2=0$ (Lissajous orbits).

We report, for the sake of completeness, the analytic expression of the coefficients
$a_x^j$ and $b_y^j$ of the inhomogeneous solution corresponding to the planetary perturbations

\begin{align}
a_x^j & =  -2\, Q \,\frac{\mu_j}{r_{0j}^2} \frac{{\cal R}_{j3} (n_{j3}^2-n_z^2+n_3^2)+ {\cal T}_{j3} \,n_3 n_{j3}}{  (n_{j3}^2+n_3^2) n_z^2+(n_3^2-n_{j3}^2)^2-2 n_z^4},\\
b_y^j & =  Q \,\frac{\mu_j}{r_{0j}^2} \frac{4  \,{\cal R}_{j3} n_3 n_{j3}+{\cal T}_{j3} (n_{j3}^2+2 n_z^2+n_3^2)}{  (n_{j3}^2+n_3^2) n_z^2+(n_3^2-n_{j3}^2)^2-2 n_z^4}.
\end{align}

\bibliography{references}

\end{document}